# Privacy by Voice: Modeling Youth Privacy-Protective Behavior in Smart Voice Assistants

Molly Campbell
*Computer Science Department*
*Vancouver Island University*
Nanaimo, Canada
molly.campbell@viu.ca

Ajay Kumar Shrestha
*Computer Science Department*
*Vancouver Island University*
Nanaimo, Canada
ajay.shrestha@viu.ca

***Abstract*—Smart Voice Assistants (SVAs) are deeply embedded in the lives of youth, yet the mechanisms driving the privacy-protective behaviors among young users remain poorly understood. This study investigates how Canadian youth (aged 16-24) negotiate privacy with SVAs by developing and testing a structural model grounded in five key constructs: perceived privacy risks (PPR), perceived benefits (PPBf), algorithmic transparency and trust (ATT), privacy self-efficacy (PSE), and privacy-protective behaviors (PPB). A cross-sectional survey of N=469 youth was analyzed using partial least squares structural equation modeling. Results reveal that PSE is the strongest predictor of PPB, while the effect of ATT on PPB is fully mediated by PSE. This identifies a critical efficacy gap, where youth's confidence must first be built up for them to act. The model confirms that PPBf directly discourages protective action, yet also indirectly fosters it by slightly boosting self-efficacy. These findings empirically validate and extend earlier qualitative work, quantifying how policy overload and hidden controls erode the self-efficacy necessary for protective action. This study contributes an evidence-based pathway from perception to action and translates it into design imperatives that empower young digital citizens without sacrificing the utility of SVAs.**

***Keywords—Privacy, Smart Device, Youth, Smart Voice Assistants, AI, Transparency, Trust, Perceived Risk, Self-Efficacy, Protective Behavior, Perceived Benefits, PLS-SEM***

## I. Introduction

Canadian youth (aged 16-24) are frequent users of Smart Voice Assistants (SVAs) such as Siri, Alexa, and Google Assistant, and show a growing reliance on voice-enabled technology [1], [2], [3]. These smart devices create privacy risks for users due to their always-listening nature, particularly in home or shared spaces where sensitive information may be inadvertently recorded [4], [5], [6]. While these devices are common in young people's lives, the mechanisms that translate youth perceptions into privacy-protective behaviors for SVAs remain underexplored.

Existing work surrounding youth privacy behaviors has often employed the privacy calculus model, examining how perceived risks and benefits shape behaviors, and has explored concepts such as the privacy paradox, trust, transparency, and self-efficacy [7], [8], [9], [10]. However, most of this work has focused on web and social media contexts, and few studies have examined these constructs together in the context of youth in SVA ecosystems [1], [11]. Even fewer studies have explicitly examined self-efficacy alongside concrete protective behaviors, such as deleting voice histories or disabling microphones.

This study extends prior focus group research with small groups of Canadian youth, which surfaced key constructs and youth-specific tensions, including always-on listening, voice logs, and parental dynamics [12]. Alongside this, a Privacy-By-Design (PbD) audit of commercial smart voice assistants revealed a variation in privacy controls and data flows across devices [13]. Together, these phases motivated a youth-centered structural model, which is tested quantitatively in the current study.

In this paper, we develop and test a five-construct structural model to examine how these factors influence youth privacy-protective behaviors in the context of SVAs. The model focuses on five key constructs informed by prior studies, literature review, qualitative work and device system audits: perceived privacy risks (PPR), perceived privacy benefits (PPBf), algorithmic transparency and trust (ATT), privacy self-efficacy (PSE), and privacy-protective behaviors (PPB). The instrument uses validated scales adapted to the SVA context, and analysis is done via Partial Least Squares Structural Equation Modeling (PLS-SEM) on an online survey of Canadian youth (N=469).

The rest of the paper is organized as follows: Section II provides background and related works. Section III contains the research model and hypotheses. Section IV holds the methodology. Section V presents the results. Section VI provides the discussion. Section VII concludes the paper.

## II. Background and Related Works

### *A. Youth, Smart Voice Assistants, and Privacy*

As young digital citizens, Canadian youth have become required to interact with smart voice assistants (SVAs) in public and home contexts [1], [2]. Everyday tasks such as setting a timer, doing a search on the web, using smart home controls, and consuming media become simple habits for those who regularly rely on these technologies [12]. The always-listening architecture of SVAs means that voices can be captured in both private and public environments, potentially turning everyday conversations into data these systems collect and process [4], [5], [6]. Although personal privacy settings can be adjusted to limit data collection on one's own device, individuals often lack the authority to manage shared household devices, where those responsible for device settings may assess risks differently.

Key privacy concerns, such as surveillance, profiling, and digital traces, have been raised by those studying youth privacy research [1], [14]. When civilians believe they are being constantly surveilled, they begin to act in ways to align with the expectations of the authority controlling the data [15]. This ultimately shapes how people speak, act, and present themselves. Confusing privacy policies and terms of use leave young digital citizens unsure about what data is stored, how long it is retained, and how it may be used [12].

Earlier focus group studies composed of 2-6 Canadian youth revealed anxieties surrounding the always-listening nature of these devices, confusion about how long data is retained and how their interactions are logged [12]. In many households, parents control the settings and permissions of the SVAs, which limits how freely youth use these devices and can lead them to feel pressured to comply with household norms, even if their privacy concerns differ from their parents' [7].

### B. *Privacy Calculus, Risk, and Benefits*

Privacy calculus theory is a theoretical framework that aims to help understand how individuals weigh the benefits and risks associated with disclosing their personal information. Privacy Calculus was first introduced by Laufer and Wolfe in 1977 as the view that privacy functions as an economic commodity that may be exchanged in return for other goods or certain advantages [16]. Armstrong and Culnan addressed the tensions that arise between the collection and use of personal information that people provide, focusing on how organizational data-handling practices play a crucial role in shaping individuals' privacy concerns [17]. Previous studies offer strong evidence-based support for the privacy calculus framework, demonstrating that individuals consistently evaluate perceived benefits against perceived risks when deciding whether to disclose personal information [7], [18], [19].

Perceived privacy risk (PPR) can be generally defined as an individual's assessment of the potential risks involved in disclosing personal information [20], [21]. While established scales like the Internet Users' Information Privacy Concerns (IUPC) have measured this construct, they have primarily been focused on web/e-commerce contexts, leaving emerging domains such as SVAs less explored [20]. Perceived risks in SVAs include factors like always-on microphones, continuous recording, and sharing with third parties [12]. Perceived privacy benefits (PPBf) refer to the expected advantages users gain from sharing personal data, such as convenience and personalized recommendations [22], [23]. Among youth, the use of these devices to simplify everyday tasks, such as sending a text, setting a reminder or playing some music, is becoming increasingly common [12]. Such interactions with SVAs have become so routine that they lead to potential privacy risks being overlooked. Both PPR and PPBf are essential in understanding user behavior in the context of SVAs and serve as key exogenous constructs in our model. Decision-making about privacy differs for youth, so it is important to study perceived risks and benefits from a youth-centered perspective [7], [8].

### C. *Transparency, Trust, and Privacy Self-Efficacy*

Trust in online systems has been tied to users' perceived clarity about what data is collected, how it is used, and by whom [24], [25]. Providing transparency regarding the algorithms used for processing, recommendations, and control mechanisms helps build trust and supports informed engagement. Privacy information for SVAs is typically distributed across several interfaces, including app settings, OS menus, account dashboards and privacy policies. A PbD technical audit of Siri, Alexa, and Google Assistant revealed that, although each device offers access to essential privacy controls and activity views, they vary in default settings, language clarity, and depth of navigation required to locate key settings [13]. These differences likely affect users' sense of transparency as well as their overall trust in the device.

Individuals vary widely in how confident they feel about finding and using privacy controls. Bandura's self-efficacy theory, first introduced in 1977, defines self-efficacy as individuals' beliefs in their capacity to plan and execute the actions required to manage specific situations [26]. Prior research in PSE repeatedly shows that individuals with higher self-efficacy are more likely to engage in proactive protective behaviors compared to those with lower self-efficacy who tend to avoid privacy settings altogether or adopt resignation-based strategies [27], [28]. These dynamics become especially apparent for youth, for whom complex settings, language and navigation can undermine their sense of competence [1], [12], [29]. Within this study, ATT will capture youths' perceptions of clarity and responsible data handling within these systems, and PSE will capture their perceived ability to act on those perceptions.

### D. *Privacy-Protective Behavior in Youth Populations*

Prior studies have documented a range of digital privacy-protective actions among youth, such as adjusting account permissions, using pseudonyms, and deleting posts or history, yet attitudes often do not translate directly into these behaviors [30], [31]. The strong contrast between young people's attitudes toward digital privacy and their behavior reveals the challenges in translating awareness into effective protective behaviors.

Youth often worry about their privacy but continue to behave inconsistently, creating what is known as the privacy paradox [32], [33], [34], in which factors such as social pressure, convenience, interface friction, and lack of awareness contribute to this inconsistency. Understanding these factors is critical for researchers seeking to empower youth to make more consistent privacy-protective choices online.

In the context of SVAs, youth privacy behaviors can include reviewing, disabling or deleting voice histories, disabling microphones, refusing certain features (such as those requiring location), or using other tools to manage their data. Quantitative research investigating these behaviors is limited and leaves gaps in our understanding of how young users navigate privacy in SVA ecosystems [12].

### E. *Summary and Research Gap*

Youth interactions with SVAs involve weighing potential risks against benefits, their trust in the system, system transparency, and their confidence in managing privacy. Despite there being extensive research on these elements individually, there is no integrated model for understanding their combined effect on youth. Based on the theoretical and empirical findings discussed, Section III presents the research model and hypotheses that guide this study.

## III. RESEARCH MODEL AND HYPOTHESES

### A. *Constructs*

Table I provides the definitions of the five key constructs examined in this study: Perceived Privacy Risk, Perceived Privacy Benefits, Algorithmic Transparency and Trust, Privacy Self-Efficacy, and Privacy Protective Behavior.

### B. *Hypotheses*

Based on the findings in Section II, the following ten research hypotheses (H1-H10) aim to examine the direct and mediated relationships between the constructs outlined in Table I. These hypotheses formalize the model and test both direct effects and underlying mechanisms, consistent with prior privacy and technology adoption research.

TABLE I. CONSTRUCTS AND DEFINITIONS

| Construct | Definition |
|---|---|
| Perceived Privacy Risk (PPR) [20], [21] | The extent to which youth feel vulnerable or at risk when using voice-activated AI apps or SVAs. |
| Perceived Privacy Benefits (PPBf) [22], [23] | The perceived advantages or conveniences gained from using SVAs that can offset privacy concerns. |
| Algorithmic Transparency and Trust (ATT) [24], [25] | The degree to which users believe that SVA developers are transparent about data practices, thereby fostering trust. |
| Privacy Self-Efficacy (PSE) [26], [35] | Users' confidence in their ability to identify, manage, and protect personal information when using SVAs. |
| Privacy-Protective Behavior (PPB) [32], [36] | Concrete actions taken by youth to safeguard their personal information and limit data collection in SVAs. |

- **H1:** PPR has a positive influence on PPB.
- **H2:** PPBf negatively influences PPB.
- **H3:** ATT positively influences PPB.
- **H4:** PSE positively influences PPB.
- **H5:** PPR negatively influences PSE.
- **H6:** PPBf positively influences PSE.
- **H7:** ATT positively influences PSE.
- **H8:** PPR has an indirect effect on PPB via PSE.
- **H9:** PPBf has an indirect effect on PPB via PSE.
- **H10:** ATT has an indirect effect on PPB via PSE.

## IV. METHODOLOGY

### *A. Study Design*

This study employs a cross-sectional online survey design as its quantitative phase, following an initial exploratory focus group and privacy device audit [12], [13]. This study received ethics approval from an institutional research ethics board. The approval reference number #103597 was given for behavioral/amendment forms, consent form, and questionnaire. The survey instruments were adapted from constructs validated in prior studies [20], [21], [22], [23], [24], [25], [26], [32], [35], [36] and tailored to the SVA context. The instruments consist of 4 indicators for each of the five constructs, PPR, PPB, PPBf, ATT, and PSE. The respective items (questions) within these constructs are detailed in Table II. We measured responses to the items on a 5-point Likert scale ranging from 1 ("Strongly disagree") to 5 ("Strongly agree"). Higher scores indicate higher levels of the underlying constructs. In addition, we collected data on control variables, including age, gender, education level, and SVA usage frequency. These variables were used for descriptive analyses and optional robustness checks but were not central to the structural model.

### *B. Participant Recruitment and Demographics*

Participants were recruited through multiple channels, including flyers, emails, personal networks, LinkedIn and through collaboration with several Canadian school districts and Universities to reach our targeted demographic of youth aged 16-24. Participation in the survey was entirely voluntary and anonymous. A monetary incentive was offered to the first 500 survey respondents, with district-specific exceptions where required. The participants were to read and accept a consent form before starting the questionnaire. By submitting the consent form, participants were indicating they understood the conditions of participation in the study as outlined in the consent form. We conducted online surveys through Microsoft Forms. Upon completing the questionnaire, participants were directed to a separate form to claim the incentive by providing their email address.

A total of 494 participants took part in the questionnaire. Responses were omitted that did not meet the demographic criteria (Canadian youths aged 16-24 with at least one SVA use in the prior month) or that contained insufficient data ($\geq$ 20% missing responses). After data cleaning, 469 valid responses remained. Remaining item-level nonresponses were left blank in the CSV and imported to SmartPLS as missing values (no imputation). SmartPLS handled the remaining missing data with its default pairwise handling during model estimation. The proportion of missing data after cleaning was low ($\leq$0.5% for any indicator). Of those 469 valid responses, 174 identified as female, 241 identified as male, 15 identified as non-binary, and 39 were missing or preferred not to say. The average age of participants was 18.65. 278 of the participants were High School students, while 183 had completed or were currently enrolled in Post-Secondary Education. The frequency of SVA varied, with 126 participants reporting daily use of SVA, 113 reporting weekly use, 38 reporting monthly use, and 190 respondents reporting they rarely used SVA devices. Table III highlights the characteristics of the demographics of the participants.

TABLE II. CONSTRUCTS AND ITEMS

| Construct | Items |
|---|---|
| Perceived Privacy Risk (PPR) | PPR1: Concern about the amount of personal information collected by SVAs.<br>PPR2: Worry about conversations being recorded without full awareness or consent.<br>PPR3: Belief that voice data could be accessed by unauthorized parties.<br>PPR4: Unease about the duration voice recordings are stored. |
| Perceived Privacy Benefits (PPBf) | PPBf1: Extent to which voice-activated assistants or SVAs save time and effort.<br>PPBf2: Worth of sharing data for the personalized features offered.<br>PPBf3: Belief that benefits outweigh data collection worries.<br>PPBf4: Appreciation for the apps learning preferences to improve services. |
| Algorithmic Transparency and Trust (ATT) | ATT1: Understanding of the types of information collected and stored.<br>ATT2: Trust that manufacturers responsibly handle voice data.<br>ATT3: Feeling that apps are upfront in explaining data processing.<br>ATT4: Belief that apps provide fair and unbiased recommendations. |
| Privacy Self-Efficacy (PSE) | PSE1: Knowledge of how to access and adjust privacy settings.<br>PSE2: Capability to prevent apps from recording when undesired.<br>PSE3: Confidence to update permissions to increase data privacy.<br>PSE4: Belief in the ability to effectively manage associated privacy risks. |
| Privacy-Protective Behavior (PPB) | PPB1: Frequency of reviewing or updating app permissions.<br>PPB2: Frequency of deleting voice search/activity history.<br>PPB3: Refusal of certain features to maintain privacy.<br>PPB4: Use of additional measures to protect data. |

TABLE III. PARTICIPANT DEMOGRAPHICS

| Characteristic | n | % |
|---|---|---|
| ***Gender*** | | |
| Blank/Missing | 4 | 0.9 |
| Female | 174 | 37.1 |
| Male | 241 | 51.4 |
| Non-binary/Other | 15 | 3.2 |
| Prefer not to say | 35 | 7.5 |
| ***Education Level*** | | |
| High School | 278 | 59.3 |
| Post-Secondary | 183 | 39 |
| Blank/Missing | 8 | 1.7 |
| ***Frequency of SVA use*** | | |
| Daily | 126 | 26.9 |
| Monthly | 38 | 8.1 |
| Rarely | 190 | 40.5 |
| Weekly | 113 | 24.1 |
| Blank/Missing | 2 | 0.4 |
| | **Mean (SD)** | |
| ***Age*** | 18.65 (2.30) | |

# V. RESULTS

For data processing, the survey responses were first coded numerically in R. Subsequently, Microsoft Excel was used to compute descriptive statistics to summarize sample characteristics. The primary analysis was conducted using a PLS-SEM approach via SmartPLS software [37]. PLS-SEM is a robust method commonly used to estimate path coefficients in structural models, widely recognized in numerous studies [7], [38], [39]. The analysis followed the established two-step procedure for SEM as suggested by [40], which involves first testing the reflective measurement models (including indicator loading, internal consistency, and convergent and discriminant validity) and then evaluating the structural model (regression analysis). We employed the path-weighting scheme in SmartPLS, which is commonly used for PLS-SEM estimation and is suited for maximizing the explained variance ($R^2$) of the endogenous constructs. Finally, we utilized the PLSpredict procedure to evaluate the model's out-of-sample predictive power.

Additionally, a nonparametric bootstrapping procedure was utilized to assess the statistical significance of the PLS-SEM results. Bootstrapping is a resampling technique that creates an empirical sampling distribution by drawing repeated samples with replacement from the original data set. For our analysis, 5,000 subsamples were generated, and a two-tailed test was conducted at a significance level of 0.05. The default path-weighting algorithm in SmartPLS was used.

## A. Descriptive Statistics

Our survey used a 5-point Likert scale to compare mean responses across five key constructs. Table IV presents the descriptive statistics (means and standard deviations) and bivariate correlation coefficients, providing initial insight into participants' perceptions. PPR had the highest mean score (M = 3.61, SD = 0.91), indicating that youth generally report a moderate to high level of concern about SVA data practices. The constructs of PPBf (M = 3.00; SD = 0.95), PSE (M = 2.97; SD = 0.83) and PPB (M = 3.03; SD = 0.77) all hovered near the scale midpoint. This suggests that participants are somewhat ambivalent about the benefits of SVA use outweighing the risks, feel moderately capable of managing their privacy concerns, and occasionally engage in protective behaviors. In contrast, ATT had the lowest mean score (M = 2.52; SD = 0.72), indicating a relatively low understanding and trust in the companies and algorithms behind voice assistants. The standard deviations for all constructs showed a reasonable spread of responses.

The correlation matrix reveals several noteworthy preliminary relationships. For instance, a higher ATT is associated with greater PPBf and higher PSE. Conversely, a stronger sense of PPR is negatively correlated with both ATT and PPBf, while being positively correlated with PPB. These patterns provide initial support for the relationships investigated in the structural model.

## B. Measurement Models

We evaluated the reflective measurement model using indicator loading analysis to assess the internal consistency, reliability and validity of the constructs.

### 1) Indicator Loading Analysis

For indicator loading analysis, we checked the factor loading of individual items, as shown in Table V, to see how each variable loaded on its own construct. The loadings greater than 0.708 are recommended, as they indicate that the construct explains more than 50% of the indicator's variance [41]. Most items loaded strongly onto their respective construct. While the loading for ATT1 (0.679) and ATT4 (0.702) fell slightly below the threshold, we retained both items as they are still well above the acceptable cutoff of 0.60 [42], and the composite reliability and Average Variance Extracted (AVE) for ATT were still satisfactory (see Table VI).

TABLE IV. DESCRIPTIVE STATISTICS

| | Mean (SD) | ATT | PPB | PPBf | PPR | PSE |
|---|---|---|---|---|---|---|
| **ATT** | 2.52 (0.718) | | | | | |
| **PPB** | 3.03 (0.767) | 0.045 | | | | |
| **PPBf** | 3.00 (0.949) | 0.414 | -0.115 | | | |
| **PPR** | 3.61 (0.909) | -0.276 | 0.307 | -0.334 | | |
| **PSE** | 2.97 (0.825) | 0.449 | 0.308 | 0.278 | -0.128 | |

TABLE V. INDICATOR LOADING

| Construct | Item | Factor Loading |
|---|---|---|
| ATT | ATT1 | 0.679 |
| | ATT2 | 0.766 |
| | ATT3 | 0.758 |
| | ATT4 | 0.702 |
| PPB | PPB1 | 0.717 |
| | PPB2 | 0.710 |
| | PPB3 | 0.709 |
| | PPB4 | 0.797 |
| PPBf | PPBf1 | 0.732 |
| | PPBf2 | 0.874 |
| | PPBf3 | 0.901 |
| | PPBf4 | 0.857 |
| PPR | PPR1 | 0.900 |
| | PPR2 | 0.878 |
| | PPR3 | 0.771 |
| | PPR4 | 0.872 |
| PSE | PSE1 | 0.714 |
| | PSE2 | 0.813 |
| | PSE3 | 0.795 |
| | PSE4 | 0.839 |

TABLE VI. CONSTRUCT RELIABILITY AND VALIDITY

| Construct | rho_a | AVE | rho_c |
|---|---|---|---|
| ATT | 0.712 | 0.529 | 0.818 |
| PPB | 0.719 | 0.539 | 0.823 |
| PPBf | 0.901 | 0.711 | 0.907 |
| PPR | 0.907 | 0.734 | 0.917 |
| PSE | 0.814 | 0.626 | 0.870 |

### 2) *Construct Reliability and Validity*

We assessed convergent validity and internal consistency reliability for each construct by calculating AVE and composite reliability metrics, as shown in Table VI. Following the guideline of [41], AVE should exceed 0.50, indicating that 50% of the variance in the items is captured by the hypothesized constructs. In our study, all constructs demonstrated good convergent validity, with AVE values exceeding the 0.50 threshold. Furthermore, both Composite Reliability (rho_c) and the more robust Dillon-Goldstein's rho (rho_a) exceeded the acceptable level of 0.70 for all constructs [41], confirming the measures' internal consistency reliability.

Discriminant validity was assessed using the heterotrait-monotrait (HTMT) ratio. As shown in Table VII, all HTMT values were well below the threshold of 0.85 [41]. The highest observed value was between PSE and ATT (0.583), indicating a moderate relationship; however, it remains acceptably distinct to confirm discriminant validity.

## C. *Structural Models*

The results of our PLS-SEM analysis are depicted in Fig. 1, featuring coefficients of determination ($R^2$), path coefficients (β), and p-values. According to Chin's guideline [43], [44], a model is considered statistically somewhat (marginally) significant (*p) with a p-value < 0.1, quite significant (**p) with a p-value < 0.01, and highly significant (***p) with a p-value < 0.001.

The model demonstrates moderate explanatory power, as seen in Fig. 1, explaining 24.2% of the variance in PPB and 24.1% of the variance in PSE. The assessment of predictive power followed a two-step process. First, all PLSpredict (10-fold cross-validation) returned positive Q²_predict values for all indicators of PSE and PPB, confirming the model's basic predictive relevance (see Table VIII).

The core of the analysis involved comparing the root mean square error (RMSE) of the PLS-SEM model against a naive linear model (LM) benchmark. For both constructs, the PLS-SEM model yielded a lower RMSE for half of the indicators. This result indicates that the model exhibits a moderate level of out-of-sample predictive performance, consistent with guidance from Shmueli et al. [45].

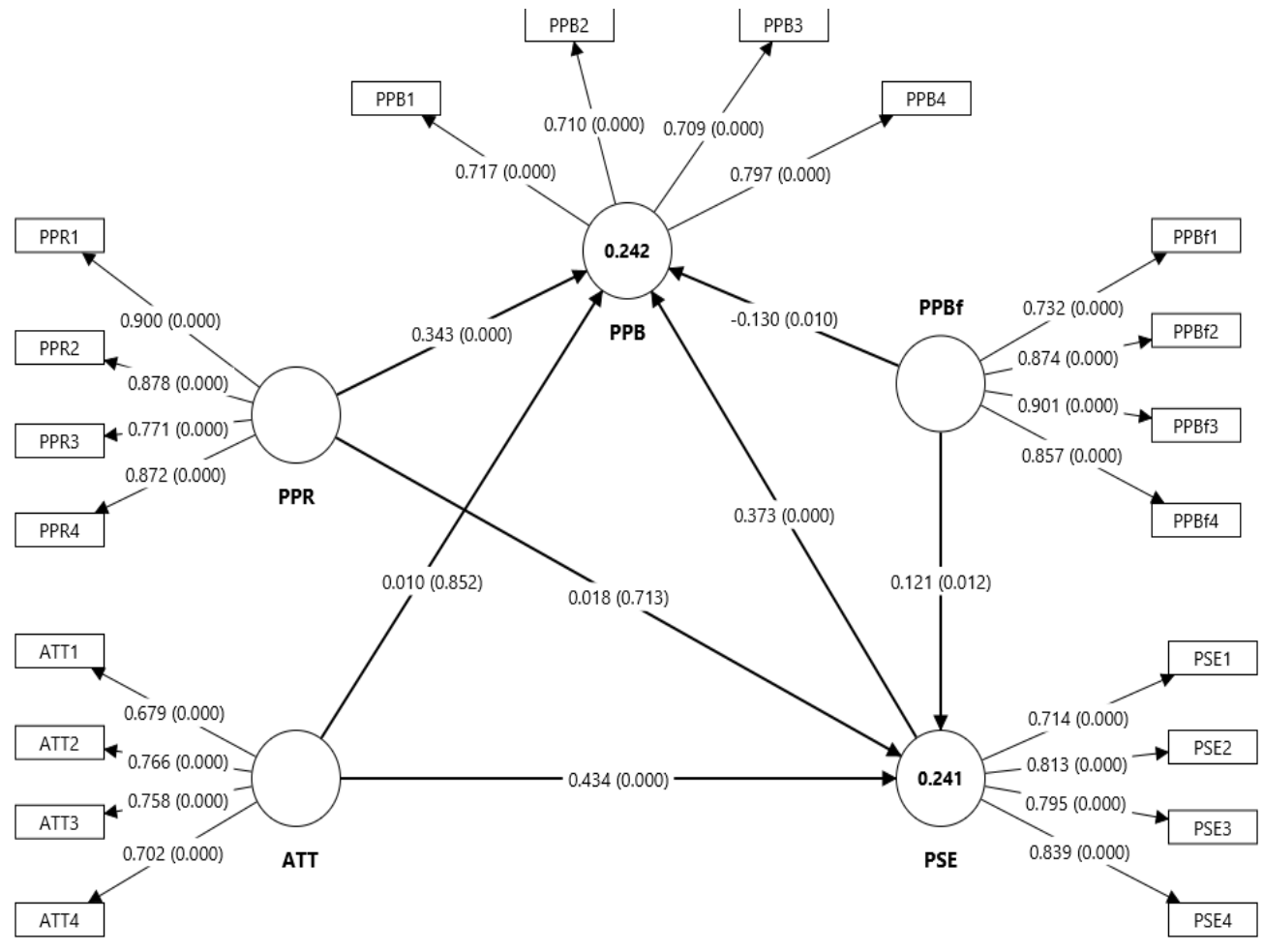

Fig. 1. PLS-SEM structural model

TABLE VII. HETEROTRAIT-MONOTRAIT RATIO

| | **ATT** | **PPB** | **PPBf** | **PPR** | **PSE** |
|---|---|---|---|---|---|
| **ATT** | | | | | |
| **PPB** | 0.227 | | | | |
| **PPBf** | 0.533 | 0.164 | | | |
| **PPR** | 0.350 | 0.394 | 0.375 | | |
| **PSE** | 0.583 | 0.404 | 0.333 | 0.165 | |

TABLE VIII. PREDICTIVE POWER

| **Item** | **Q2_predict** | **RMSE** | |
|---|---|---|---|
| | | ***PLS-SEM*** | ***Linear Model*** |
| PSE1 | 0.050 | 1.031 | 0.978 |
| PSE2 | 0.150 | 1.006 | 1.003 |
| PSE3 | 0.112 | 0.931 | 0.939 |
| PSE4 | 0.236 | 0.898 | 0.902 |
| PPB1 | 0.065 | 1.022 | 1.012 |
| PPB2 | 0.005 | 1.073 | 1.075 |
| PPB3 | 0.085 | 0.867 | 0.852 |
| PPB4 | 0.079 | 1.062 | 1.072 |

### 1) *Direct Effect Analysis*

To ensure the robustness of our direct effect results, we assessed the variance inflation factor (VIF) to check for collinearity. In line with the guideline that VIF values be ideally < 3, all our values fell well below this threshold (maximum VIF = 1.47), indicating no concern for multicollinearity in our model [41] (see Table IX for results). The analysis of direct effects, as seen in Table IX, revealed several significant relationships. Specifically, the paths from PPR to PPB (β = 0.343; p < 0.001), PSE to PPB (β = 0.373; p < 0.001), PPBf to PSE (β = 0.121; p < 0.1), and ATT to PSE (β = 0.434; p < 0.001) were positive and significant, providing support for H1, H4, H6, and H7. The relationship between PPBf and PPB (β = -0.130; p < 0.1) was negative and significant, thus supporting H2. In contrast, the path between ATT and PPB (β = 0.010; p > 0.1), and PPR to PSE (β = 0.018; p > 0.1) were not significant, leading to the rejection of H3 and H5. The effect size ($f^2$) of each path was calculated to determine its impact and followed conventional benchmarks (0.02 small, 0.15 medium, 0.35 large). The paths from PPR to PPB ($f^2$ = 0.134) and PSE to PPB ($f^2$ = 0.139) demonstrated a small-to-medium effect size. The path from ATT to PSE ($f^2$ = 0.204) showed a medium effect, while the paths from PPBf ($f^2$ = 0.017) and PPBf to PSE ($f^2$ = 0.015) demonstrated small (near the threshold) effects.

### 2) *Mediation Analysis*

To test hypotheses H8-H10, we conducted an indirect (mediation) effects analysis using bootstrap-based 95% confidence intervals (5,000 resamples). Mediation was considered statistically supported if the 95% CI did not include zero. The analysis, as seen in Table X, revealed that the indirect path from PPR to PPB through PSE was not significant (β = 0.007; p > 0.1), leading to the rejection of H8. In contrast, the indirect effect of PPBf on PPB through PSE was positive and significant (β = 0.045; p < 0.1), supporting H9. Since the direct effect from PPBf to PPB remains significant (β = -0.130; p < 0.1), this indicates that PSE partially mediates the relationship. Finally, the indirect path from ATT to PPB through PSE was positive and highly significant (β = 0.162; p < 0.001), supporting H10. Given that the direct effect from ATT to PPB was not significant (β = 0.010; p > 0.1), this shows that PSE fully mediates the relationship between ATT and PPB.

TABLE IX. DIRECT EFFECT ANALYSIS

| Structural path | Std β | T | P | $f^2$ | VIF |
|---|---|---|---|---|---|
| PPR → PPB | 0.343 | 7.096 | 0.000 | 0.134 | 1.158 |
| PPBf → PPB | -0.130 | 2.574 | 0.010 | 0.017 | 1.310 |
| ATT → PPB | 0.010 | 0.187 | 0.852 | 0.000 | 1.466 |
| PSE → PPB | 0.373 | 7.869 | 0.000 | 0.139 | 1.317 |
| PPR → PSE | 0.018 | 0.367 | 0.713 | 0.000 | 1.157 |
| PPBf → PSE | 0.121 | 2.503 | 0.012 | 0.015 | 1.290 |
| ATT → PSE | 0.434 | 10.160 | 0.000 | 0.204 | 1.218 |

TABLE X. INDIRECT EFFECT ANALYSIS

| Structural path | Std β | P | 95% CI |
|---|---|---|---|
| PPR → PSE → PPB | 0.007 | 0.719 | [-0.028, 0.044] |
| PPBf → PSE → PPB | 0.045 | 0.019 | [0.009, 0.085] |
| ATT → PSE → PPB | 0.162 | 0.000 | [0.114, 0.221] |

## VI. DISCUSSION

The goal of our research is to investigate the complex mechanisms driving youth privacy behaviors with SVAs. By integrating quantitative structural modeling with prior qualitative insights [12] and a PbD audit [13], we move beyond simply identifying the privacy paradox to explaining its underlying pathways. Our findings confirm that youth engagement with SVAs is a dynamic trade-off, but one that is critically mediated by a young person's belief in their own ability to manage their privacy.

### A. *The Importance of Self-Efficacy*

Our model revealed several compelling narratives. The most powerful driver of PPB is self-efficacy. This highlights the importance of PSE; without the confidence to act, even high privacy concerns fail to translate into tangible protective measures, pointing to a crucial need for privacy education that builds practical skills.

The relationship between perceived benefits and behavior is nuanced. While benefits directly discouraged protective behavior, as predicted by the privacy calculus, they also had a small but positive effect on protective behavior when mediated by self-efficacy. This suggests that when youth find value in an SVA, it can boost their confidence in using it, which can foster protective actions.

Most revealing is that the relationship between ATT and behavior is fully mediated by PSE. Our analysis revealed that ATT had no significant effect on PPB, but a strong and positive effect on PSE. This indicates that a sense of trust and transparency in device design does not automatically lead to protective behavior. This trust empowers users, and it's this empowerment that leads to action.

### B. *Refining the Privacy Calculus for Youth*

These findings help in refining established privacy theories for the youth SVA context. The Privacy Calculus is validated but complicated. As seen with the direct negative effect that PPBf has on PPB, this confirms that youth do trade privacy for convenience. However, the positive indirect effect of perceived benefit on protective behaviors via self-efficacy reveals that this trade-off is not binary. Convenience may be a gateway to engagement that can build the confidence needed for action.

Our model provides a strong explanation for the Privacy Paradox. The paradox, where concerns fail to translate into action, is not just apathy but an efficacy gap. We see that PPR directly motivates PPB, showing that concerned youth want to act. However, the non-significant path from PPR to PSE indicates that feelings of risk do not translate to a sense of control. Without accessible tools and clear guidance, great concern can lead to privacy fatigue and hopelessness rather than effective action.

### C. *Convergence with Qualitative Themes*

Our quantitative model reinforces the themes identified in prior qualitative focus group research [12]. Participants reported "always-listening" anxiety and confusion about data logs is reflected in the reported high PPR and low ATT scores. The strongest relationship in our model, from PSE to PPB, directly quantified a central reported struggle. The theme of "Low Navigation Efficacy", where participants reported feeling confused about how to navigate privacy settings, is the embodiment of low PSE. Our model confirms that this lack of confidence is the biggest barrier to protective action. The significant and positive relationship between PPR and PPB captures the motivation behind the extreme physical mitigation reported, such as physically disconnecting a microphone. High risk perception directly drives protective behavior, even drastic ones. The significant relationship of ATT to PSE quantifies a key qualitative insight: transparency is the foundation of trust, and trust is a prerequisite for self-efficacy. The qualitative findings that policy overload and unclear data retention practices reduce trust are seen in this relationship. The strong ATT to PSE pathway confirms that for many youth, opaque data practices directly corrode the confidence needed to act. The small but significant positive indirect effect of PPBf on PPB via PSE aligns with the finding that efficacy is device-conditional. Familiarity and utility on one's primary device can foster the PSE needed to take protective action.

### D. *Connection to Privacy-by-Design Audit*

A prior PbD audit [13] provides a tangible explanation for the relationships observed in this analysis. The audit reports rubric-based scores and retention as time-to-verified effect. Differences across commercial smart voice assistants, such as Siri, Alexa, and Google designs directly influence the ATT and PSE constructs in our model. Google Home's high usability score and real-time feedback are design features that should directly enhance PSE by making the outcomes of actions clear and immediate. Siri's high compliance score and strong consent mechanisms exemplify high ATT, which our model shows should boost PSE. However, its fragmented settings menu, which confused participants in UX testing, creates a navigational barrier that undermines self-efficacy; this aligns with the indirect effect of ATT on PPB. Alexa's intuitive navigation across platforms, which supports PSE, is hindered by its opaque data retention policies, which lower users' trust. The finding that disabling voice history can take up to 36 hours erodes trust and confirms the qualitative theme of "retention unknowns", directly influencing the link between a user's action and their perceived control.

### E. *Design Implications for Smart Voice Assistants*

The central role of PSE as a critical mediator between perception and action provides a clear mandate for SVA designers. The goal must be to build systems that empower youth. Our findings point to several concrete design priorities.

*1) Design for Privacy Self-Efficacy:* Interfaces must be designed to make users feel capable. This requires creating a unified privacy hub with single-jump access to all key controls, eliminating the confusion that lowers navigation efficacy. Manufacturers should replace legal jargon with plain-language, action-oriented explanations and embed device-conditional, 30-second micro-tutorials directly within settings to guide users through key tasks.

*2) Implement Practical, Local Transparency:* Trust is built through verifiable actions, not just policies. Systems should provide just-in-time prompts that explain why specific data is needed at the moment of request. To combat retention opacity, interfaces must offer readable activity logs and audit trails, allowing users to see what is being recorded and verify that deletion has occurred. Clear, persistent status indicators should extend to data retention states (e.g., listening, storing, deletion).

*3) Configure Defaults to Combat Privacy Fatigue:* To counteract the helplessness that leads to privacy fatigue, systems should do the heavy lifting for the users. This involves setting conservative, youth-oriented defaults, such as auto-deletion of voice history after 30 days and opt-out of personalized advertising. Implementing low-friction privacy check-ups that prompt users with simple, pre-set "privacy profiles" (e.g., "More Private" vs. "More Personalized") can allow for easy resets.

*4) Acknowledge the Youth-Family Context:* SVAs are often shared devices, requiring designs that acknowledge a social context. Platforms should implement household multi-user modes. For younger users, providing simplified, tiered privacy dashboards and guardian oversight for critical settings can offer age-appropriate protection, while embedded educational prompts can build digital literacy in context.

### F. Limitations

This study has some limitations; the cross-sectional design only offers a snapshot in time, constraining our ability to make causal inferences and capture the evolving nature of privacy attitudes and behaviors. Future research could benefit from employing random sampling methods and longitudinal designs to validate these results. Furthermore, all constructs are measured through self-reported data, which is susceptible to social desirability bias and recall issues. We did not collect actual behavioral observation data to corroborate reported privacy protective behaviors. Finally, while our recruitment successfully targeted Canadian youth (aged 16-24) with recent SVA experience, the anonymous nature of the survey meant we did not collect detailed demographic data. Consequently, our sample may not be fully representative of the demographic diversity within the population, and the findings should be validated across a more balanced and diverse range of youth in future studies.

## VII. Conclusion

This research moves beyond simply identifying the privacy paradox to explaining its underlying mechanisms in the context of youth and SVAs. By integrating a quantitative structural model with prior qualitative insights and a PbD audit, this study demonstrates that youth engagement is not a simple trade-off between risk and benefit, but a pathway mediated by self-efficacy. We have shown that the most significant driver of protective behavior is a youth's confidence in their ability to act. While perceived risks motivate action and perceived benefits discourage it, it is the empowerment created by transparency and trust that ultimately enables it. These findings provide a clear, evidence-based mandate for manufacturers, policymakers, and educators. To empower young digital citizens, the focus must shift beyond merely informing users to actively enabling them. This involves embedding intuitive privacy controls, providing verifiable transparency, and implementing conservative defaults that protect users by design. By adopting the design implications outlined, we can create a digital ecosystem where convenience does not come at the cost of control. Future research should build upon this qualitative and quantitative foundation by co-designing and empirically evaluating the proposed privacy features with youth. Furthermore, longitudinal studies are needed to track how these relationships evolve over time. This study offers a path toward SVA governance and design that respects youth as capable digital citizens, ensuring the future of AI voice technology is both convenient and empowering.

## Acknowledgment

This project has been funded by the Office of the Privacy Commissioner of Canada (OPC); the views expressed herein are those of the authors and do not necessarily reflect those of the OPC.